\documentclass[useAMS,usenatbib]{mnras}

\usepackage[a4paper]{geometry}
\skip\footins=19.5pt plus 12.0pt minus 1.0pt
\usepackage{amsmath}
\usepackage[english]{babel}
\usepackage[T1]{fontenc}
\usepackage{verbatim}
\usepackage{graphicx,subfigure,float}
\usepackage{txfonts}
\usepackage{upgreek}

\usepackage[all]{hypcap} 
\usepackage{multirow}

\usepackage{etoolbox}
\makeatletter
\patchcmd\@combinedblfloats{\box\@outputbox}{\unvbox\@outputbox}{}{%
   \errmessage{\noexpand\@combinedblfloats could not be patched}%
}%
 \makeatother

\def\hi{\ifmmode{\rm HI}\else{H\/{\sc i}}\fi} 
\def\ha{\ifmmode{{\rm H}\upalpha}\else{H$\upalpha$}\fi}

\newcommand {\de}{^{\circ}}
\newcommand {\mo}{{\rm M}_\odot}

\newcommand {\moyr}{\,{\rm M_\odot\,\rm yr}^{-1}}

\newcommand{\bba}{$^{\scriptstyle 3\mathrm{D}}$B{\sc arolo}}

\newcommand{\vir}[1]{``#1''}
\newcommand {\vrot}{{V_\mathrm{rot}}}
\newcommand {\vflat}{{V_\mathrm{flat}}}
\newcommand {\vdisp}{{\sigma_\mathrm{gas}}}
\newcommand{\vsigmaratio}{V / \sigma}
\newcommand{\gal}{Sp1149}
\newcommand{\hiz}{high-$z$}
\newcommand{\pmo}[2]{^{+#1}_{-#2}}
\newcommand{\ang}{$\mathrm{\AA}$}
\newcommand{\oii}{$\mathrm{[O\,\textsc{ii}]}$}
\newcommand{\oiii}{$\mathrm{[O\,\textsc{iii}]}$}
\newcommand{\nii}{$\mathrm{[N\,\textsc{ii}]}$}
\newcommand{\refsec}[1]{Section \ref{#1}}

\title[Kinematics of the SN Refsdal host galaxy]{Kinematics of the SN Refsdal host revealed by MUSE: \\a regularly rotating spiral galaxy at z $\simeq$ 1.5}
 
\author[Di Teodoro et al.]{
  E.~M.~Di Teodoro$^{1}$\thanks{E-mail: enrico.diteodoro@anu.edu.au},
  C.~Grillo$^{2,3}$,
  F.~Fraternali$^{4,5}$,
  R.~Gobat$^{6}$,
  W.~Karman$^{4}$, 
  \newauthor 
  A.~Mercurio$^{7}$,
  P.~Rosati$^{8}$,
  I.~Balestra$^{9}$,
  G.~B.~Caminha$^{8}$,
  K.~I.~Caputi$^{4}$,
  \newauthor 
  M.~Lombardi$^{2}$,
  S.~H.~Suyu$^{10,11,12}$, 
  T.~Treu$^{13}$, and
  E.~Vanzella$^{14}$ \\
  $^{1}$Research School of Astronomy and Astrophysics - The Australian National University, Canberra, ACT, 2611, Australia \\
  $^{2}$Dipartimento di Fisica, Universit\`a  degli Studi di Milano, via Celoria 16, I-20133 Milano, Italy\\
  $^{3}$Dark Cosmology Centre, Niels Bohr Institute, University of Copenhagen, Juliane Maries Vej 30, DK-2100 Copenhagen, Denmark\\
  $^{4}$University of Groningen, Kapteyn Astronomical Institute, Postbus 800, 9700 AV Groningen, The Netherlands\\
  $^{5}$Dipartimento di Fisica e Astronomia, Universit\`a di Bologna, 6/2, Viale Berti Pichat, 40127 Bologna, Italy\\
  $^{6}$School of Physics, Korea Institute for Advanced Study, Hoegiro 85, Dongdaemun-gu, Seoul 02455, Korea\\
  $^{7}$INAF - Osservatorio Astronomico di Capodimonte, Via Moiariello 16, I-80131 Napoli, Italy\\
  $^{8}$Dipartimento di Fisica e Scienze della Terra, Universit\`a degli Studi di Ferrara, Via Saragat 1, I-44122 Ferrara, Italy\\
  $^{9}$University Observatory Munich, Scheinerstrasse 1, D-81679 Munich, Germany\\
  $^{10}$Max-Planck-Institut f\"ur Astrophysik, Karl-Schwarzschild-Str. 1, D-85748 Garching, Germany\\
  $^{11}$Institute of Astronomy and Astrophysics, Academia Sinica, P.O. Box 23-141, Taipei 10617, Taiwan\\
  $^{12}$Physik-Department, Technische Universit\"at M\"unchen, James-Franck-Stra\ss{}e~1, 85748 Garching, Germany\\
  $^{13}$Department of Physics and Astronomy, University of California, Los Angeles, CA, USA 90095-1547\\
  $^{14}$INAF - Osservatorio Astronomico di Bologna, via Ranzani 1, I-40127 Bologna, Italy
 }

\begin{document}

\date{}

\pagerange{\pageref{firstpage}--\pageref{lastpage}} \pubyear{2018}

\maketitle
  
\begin{abstract}
We use Multi Unit Spectroscopic Explorer (MUSE) observations of the galaxy cluster MACS J1149.5+2223 to explore the kinematics of the grand-design spiral galaxy (\gal) hosting the SN \vir{Refsdal}.  \gal\ lies at $z\simeq1.49$, has a stellar mass $M_*\simeq5\times10^9 \, \mo$, a star-formation rate $\mathrm{SFR} \simeq1-6 \, \moyr$ and represents a likely progenitor of a Milky-Way-like galaxy.
All the four multiple images of \gal\ in our data show strong \oii-line emissions pointing to a clear rotation pattern. We take advantage of the gravitational lensing magnification effect ($\simeq 4 \times$) on the \oii\ emission of the least distorted image to fit 3D kinematic models to the MUSE data-cube and derive the rotation curve and the velocity dispersion profile of \gal. 
We find that the rotation curve steeply rises, peaks at $R\simeq1$ kpc and then (initially) declines and flattens to an average $\vflat = 128\pmo{29}{19}$ km/s. The shape of the rotation curve is well determined but the actual value of $\vflat$ is quite uncertain because of the nearly face-on configuration of the galaxy. 
The intrinsic velocity dispersion due to gas turbulence is almost constant across the entire disc with an average of $27\pm5$ km/s. 
This value is consistent with $z=0$ measurements in the ionized gas component and a factor 2-4 lower than other estimates in different galaxies at similar redshifts.
The average stellar-to-total mass fraction is of the order of one fifth.
Our kinematic analysis returns the picture of a regular star-forming, mildly turbulent, rotation-dominated ($\vsigmaratio\simeq5$) spiral galaxy in a 4 Gyr old Universe.
\end{abstract}


\section{Introduction}

MACS J1149.5+2223 \citep[hereafter MACS1149,][]{Ebeling+07} is a well-known galaxy cluster at $z = 0.542$. It has been targeted by several observational programs, including the Cluster Lensing And Supernova survey with Hubble \citep[CLASH,][]{Postman+12}, the Hubble Frontier Fields \citep[HFF,][]{Lotz+14} and the Grism Lens-Amplified Survey from Space \citep[GLASS,][]{Schmidt+14,Treu+15} with the Hubble Space Telescope (HST).
MACS1149 shows numerous strong-lensing features, like multiply imaged galaxies and elongated arcs, and lensing models have been developed in several studies \citep[e.g.,][]{Smith+09, Zitrin+11,Johnson+14,Oguri15,Grillo+16,Jauzac+16}. 
This cluster has also revealed a magnified image of one of the youngest-known galaxies at $z\simeq9.6$ \citep{Zheng+12} and the first case of a multiply-imaged, spatially-resolved supernova (SN), commonly referred to as SN \vir{Refsdal} \citep{Refsdal64}, in a background magnified spiral galaxy \citep{Kelly+15, Kelly+16, Rodney+16, Treu+16}.

\begin{figure*}
\label{fig:maps}
\center
\includegraphics[width=\textwidth]{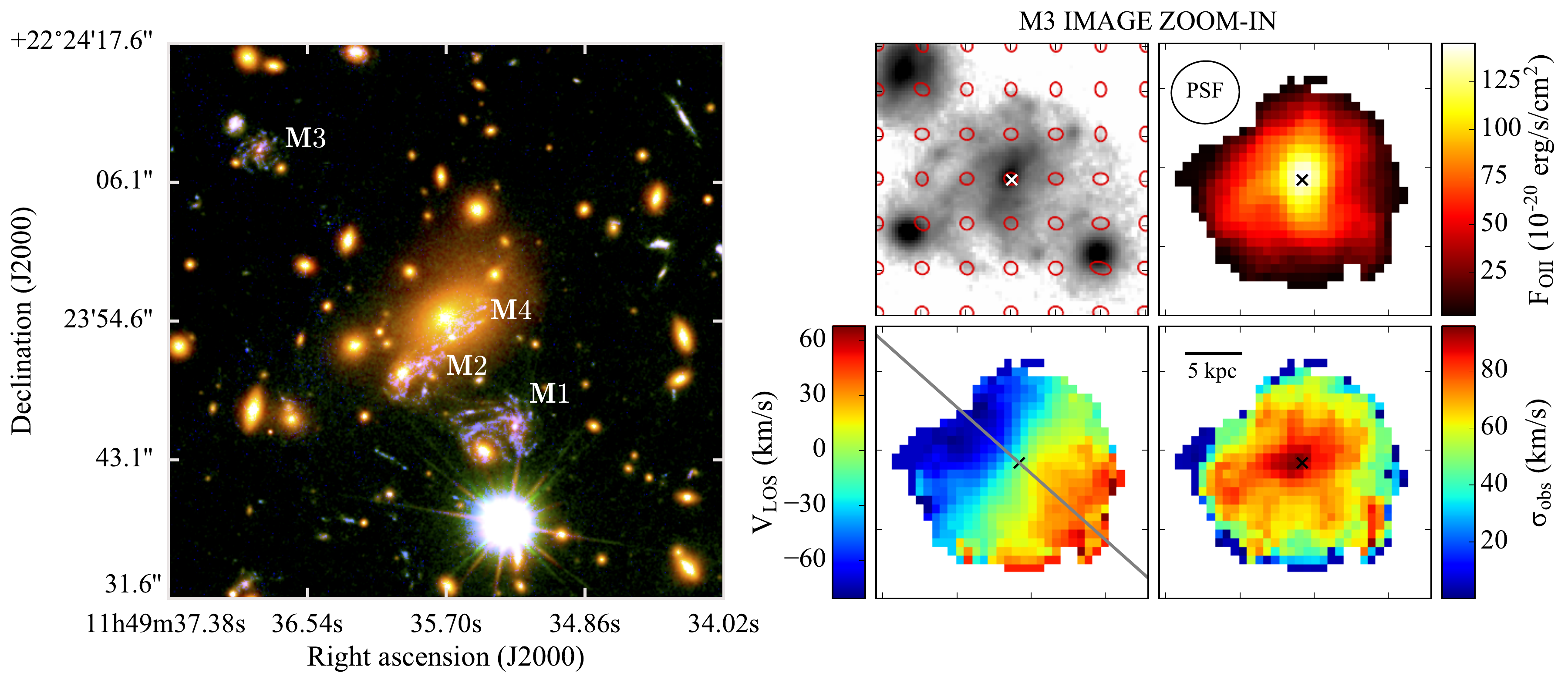}
\caption{\emph{Left panel}: RGB image of the innermost 150 kpc (in radius) of the cluster MACS J1149.5+2223 at $z=0.542$. The image has been obtained from HST data from the CLASH survey in the F150W/F814W/F225W wide filters for the RGB channels, respectively. Four multiple images of the magnified spiral galaxy \gal\ at $z=1.49$ are easily identifiable and labelled M1-M4. \emph{Right panels}: zoom-ins of the M3 image of \gal\ used in this work. We show the F814W HST image from CLASH (\emph{top-left}), the total \oii\ intensity map (\emph{top-right}), the velocity field (\emph{bottom-left}) and the observed velocity dispersion field (i.e. uncorrected for observational biases, \emph{bottom-right}). Boxes are sized $7''\times7''$.
Red ellipses on the HST image represent the gravitational lensing magnification and distortion (see text).
Moment maps are derived from our MUSE data-cube by fitting a double Gaussian function to each spatial pixel spectrum. 
For a better visualization, these maps were extracted after spatially smoothing the original data-cube with a box kernel of 3x3 pixels.
Part of \gal\ in the MUSE maps is cut (top-left corner) because of the foreground galaxy visible in the HST image. 
The size of the PSF of the MUSE observations is shown on the intensity map. White/black crosses indicates the adopted centre of the galaxy, the gray straight-line on the velocity field denotes the estimated position angle. Moment maps are only displayed for educational purposes and are not used in our kinematic analysis.}
\vspace*{10pt}
\end{figure*}

The SN Refsdal exploded in a spiral galaxy (\gal, hereinafter) at $z=1.49$ behind MACS1149. Thanks to the lensing magnification effect provided by the cluster, we can study this spiral galaxy with a level of detail comparable to that reached at $z = 0.1$ \citep{Smith+09, Karman+16}. 
The galaxy reveals itself in four (two full and two partial) multiple images (\autoref{fig:maps}, \emph{lefthand panel}) that are highly magnified \citep[a factor $\mu$ = 4 - 20, depending on the image, e.g.][]{Zitrin&Broadhurst09, Grillo+16} and has been studied in several works well before the appearance of the SN Refsdal. 
\gal\ is a nearly face-on disc-dominated galaxy with a stellar mass $M_* \simeq 10^{9.5-9.7}\,\mo$ and bulge-to-total light ratio $\mathrm{B/T}\simeq0.4-0.5$ \citep[rest-frame B-band,][]{Smith+09}, showing at least four extended and prominent spiral arms that host several star-forming regions sized 50-100 pc \citep{Adamo+13}. 
Rest-frame UV \citep[V$_{555}$ band,][]{Smith+09} and $\ha$ observations \citep{Livermore+12} revealed a normal star-formation rate (SFR) of $1-6 \,\moyr$. 
\gal\ almost perfectly lies on the galaxy main sequence (i.e. SFR-M$_*$ relation) at $1<z<2$ \citep[e.g.][]{Speagle+14} and represents a likely progenitor of a Milky Way-like galaxy in the local Universe \citep{Behroozi+13,vanDokkum+13}.

The large magnification  factor allows us to trace the kinematics of \gal\ with an accuracy otherwise unachievable for a galaxy at $z\simeq1.5$. The kinematic properties of young star-forming galaxies are still controversial, although the number of \hiz\ studies has been quickly increasing in the last decade \citep[e.g.][]{ForsterSchreiber09, Gnerucci+11, Contini+12, Wisnioski+15, Stott+16}. 
The current understanding is that \hiz\ galaxies are settled in turbulent disky structures, namely marginally-stable rotating discs with a significant contribution of random motions to the dynamical support of the system \citep[see, for example, the review of][]{Glazebrook13}. 
The mechanisms that generate such a high turbulence (i.e. high velocity dispersion) are still debated \citep[e.g.,][]{Green+10,Swinbank+12} and it is not clear how \hiz\ galaxies evolve into the lowly-turbulent and rotation-dominated systems that we observe in the local Universe \citep[e.g.,][]{Bournaud+09}. 
In general, estimating the intrinsic kinematics of \hiz\ galaxies is challenging because observations have very low spatial resolution and signal-to-noise ratios (S/N).
The low spatial resolution makes it difficult to derive meaningful kinematic maps and disentangle rotation velocity from velocity dispersion because of the well-known \vir{beam smearing} effect \citep[e.g.,][]{Bosma78, Epinat+10}.
In particular, the high values of velocity dispersion derived for many high-$z$ galaxies could be, at least partially, due to an instrumental bias (line broadening) and not to an intrinsically high velocity dispersion of the gas \citep[see e.g.,][]{Davies+11,DiTeodoro+16}.
In this context, gravitational lensing helps to improve both the S/N and the linear spatial resolution through magnification \citep[e.g.][]{Jones+10a, Newman+15,Leethochawalit+16,Yuan+17}, minimizing the impact of beam smearing.

The former attempts to derive the kinematics of \gal\ returned conflicting results. \cite{Yuan+11} used the integral field spectrograph (IFS) OSIRIS \citep{Larkin+06} on the Keck Telescope to trace the \ha-\nii\ emission in one of the multiple images (M1 in \autoref{fig:maps}) of \gal\ and found a maximum rotation velocity $V_\mathrm{max}=210$ km/s (assuming an inclination of $45\de$), with a rotation-to-dispersion ratio $\vsigmaratio \simeq 4$. The same OSIRIS observations led instead \cite{Livermore+15} to estimate a $V_{2.2} = 59$ km/s with a $\vsigmaratio < 2$, where $V_{2.2}$ is the rotation velocity at 2.2 times the scale length of the exponential disc. 
These studies were however not targeted at accurately defining the kinematics of \gal\ and their results could be biased by the very low S/N of OSIRIS observations, the non-negligible distortion of the M1 image due to gravitational lensing and an improper account for beam smearing. 
In a more recent paper, \citet{Mason+16} presented the first results of the KMOS Lens-Amplified Spectroscopic Survey (KLASS), a survey with the KMOS IFS \citep{Sharples+13} aimed at investigating the kinematics of lensed galaxies at $1 \lesssim z \lesssim 3$, including \gal. 
They have \oiii\ observations of both the M1 and M3 images: their 3D modelling of M1 returned a  $V_\mathrm{max} = 227$ km/s with an inclination of $12\de$ and a low gas velocity dispersion $\sigma = 15$ km/s ($\vsigmaratio \simeq 15$). M3 is not spatially resolved in their data and they just set an upper limit to the velocity dispersion of about $50$ km/s.

In this paper we try to nail down the kinematic properties of \gal\ by using high-quality IFS data from the Multi Unit Spectroscopic Explorer \citep[MUSE]{Bacon+10} on the Very Large Telescope (VLT). 
We modelled the \oii\ data-cube of the least magnified and least distorted image of \gal\ (M3 in \autoref{fig:maps}) using an updated version of the \bba\ code \citep{DiTeodoro&Fraternali15}. 
Unlike other 2D techniques based on the modelling of kinematic maps, \bba\ works directly in the 3D observational space and is essentially unaffected by beam smearing issues. 
This means that, as long as a galaxy is resolved, the spatial resolution does not significantly affect the derived kinematic properties.
Three-dimensional fitting codes, such as \bba\ or GalPaK$^\mathrm{3D}$ \citep{Bouche+15}, currently provide the most advanced kinematic modelling of low spatial resolution observations \citep[see e.g.][]{Bacon+15, Schroetter+15, Contini+16}. 

The paper is organized as follows. In \refsec{sec:data} we summarize the MUSE observations used in this work. 
\refsec{sec:stellarmass} and \refsec{sec:kinematics} describe the stellar mass estimates and the kinematic modelling procedure, respectively. 
We show and discuss our findings in \refsec{sec:results}, comparing them to those published in the literature in \refsec{sec:comp} and drawing our conclusions in \refsec{sec:summary}.
Throughout this work, we use a flat $\Lambda$CDM cosmology with $\Omega_\mathrm{m,0} = 0.27$, $\Omega_{\Lambda,0}=0.73$ and $H_0 = 70$ km s$^{-1}$ Mpc$^{-1}$. In this cosmology framework, 1 arcsec corresponds to 8.7 kpc and the look-back time is 9.5 Gyr at $z = 1.5$.

\section{Data}
\label{sec:data}
MACS1149 was observed with MUSE at the VLT under the program ID 294.A-5032 (P.I.: C. Grillo). 
The total integration time is 4.8 hours, divided in 12 exposures of 24 minutes each. 
Ten exposures are taken under a seeing lower than 1.1$''$, two exposures have worse observational conditions and a seeing of about 2$''$.
Observations are centered at coordinates $\upalpha$ = 11:49:35.75, $\updelta$ = +22:23:52.4 (J2000) and have one single pointing. 
MUSE guarantees a wide field of view (FOV) of 1 arcmin$^2$, meaning that the four multiple images of \gal\ lie all inside our data. In \autoref{fig:maps} (left panel), we show a composite RGB image of the central regions of MACS1149 \citep[CLASH survey,][]{Postman+12}. The image extends about two thirds of the MUSE FOV. The four images of \gal, labelled M1-M4, clearly stand out thanks to the blue-UV light from the young stellar populations in the spiral arms.

For a comprehensive description of MUSE observations and data reduction process we refer to \citet{Grillo+16} and \citet{Karman+15}, respectively. Our final data-cube has a pixel scale of 0.2$''$ and a spectral channel width of 1.25 \ang. The whole spectroscopic coverage of data spans from 4750 \ang\ to 9350 \ang, corresponding to a rest-frame range $\sim$1910 - 3750 \ang\ for \gal\ at $z$ = 1.49. The \oii\  $\uplambda\uplambda$3726-3729\ang\ doublet represents the strongest emission feature from the ionized gas component in this range. 
All the four multiple images of \gal\ in our data show strong \oii\ emission lines. Out of four multiple \oii\ detections of \gal, M2 and M4 are severely distorted and not ideal to be kinematically modelled before a complicated reconstruction on the source plane.
M1 is highly magnified \citep[$\mu\simeq20$, e.g.][]{Smith+09} and less stretched than M2/M4 and it has been used in previous IFS studies to infer some kinematic information \citep{Yuan+11,Livermore+15,Mason+16}. 
The western region of this image is however affected by a secondary lensing effect due to the proximity of an elliptical galaxy member of the cluster (see M1 in \autoref{fig:maps}). 

The last image, M3, lies at $\sim 20$ arcsec from the central core of the cluster and it is the least magnified one, with an almost isotropic and constant distortion over its entire surface \citep[e.g.,][]{Zitrin&Broadhurst09}. 
In the HST zoom-in of the M3 image (\autoref{fig:maps}, \emph{top-left} map in the \emph{right panels}), we show the gravitational lensing effect in terms of magnification and stretching predicted by the best-fit MLV-G12F median model by \citet{Grillo+16} with {\sc Glee}\footnote{Lens modelling software developed by A.~Halkola and S.~H.~Suyu \citep{Suyu&Halkola10, Suyu+12}.}.
Red ellipses represent how a circular source with radius $r$ = 1 pixel = 0.065$''$ is lensed in different regions of the M3 image. 
The magnification factor is $\mu\sim4$, which means that ellipses have semi-major axes $a\approx r\mu^{0.5} $. 
The ellipticity of the ellipses exemplifies the anisotropic distortion effect, which is very small and does not significantly vary across the \gal\ disc.
The M3 image offers therefore the favourable chance of studying the kinematics of a magnified galaxy without the need of reconstructing the image on the source plane. 
We stress that, thanks to full 3D modelling of the data-cube, the loss of magnification with respect to the other images does not reduce our capability of deriving a reliable kinematics. 

\begin{figure*}
\label{fig:chmaps}
\center
\includegraphics[width=\textwidth]{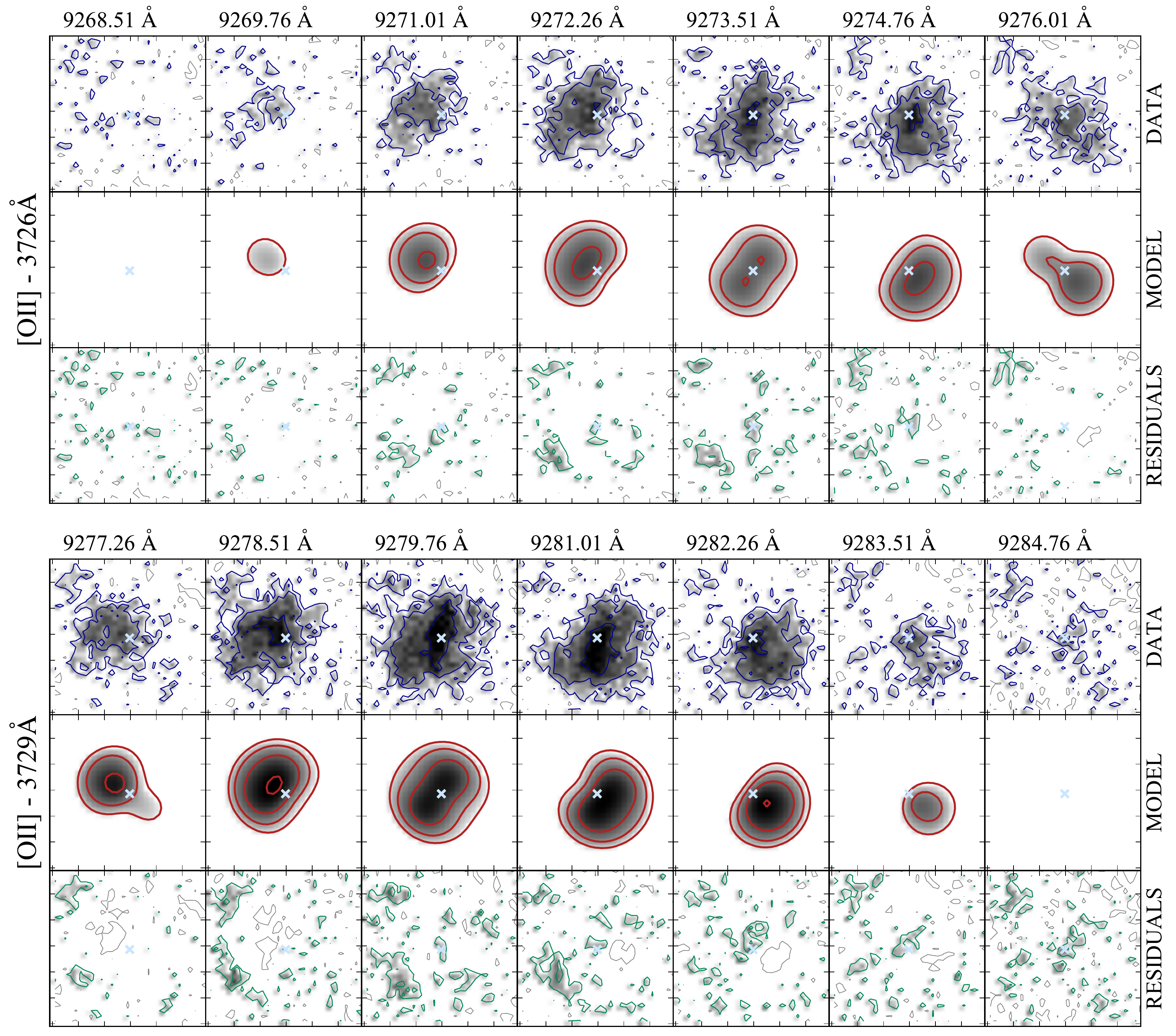}
\caption{Comparison between the M3 MUSE data-cube of \gal\ (blue contours) and the best-fit \bba\ model (red contours). Green contours represent the data-minus-model residuals. 
The plots show all the wavelengths (channel maps) where the \oii\ emission is present. 
The spectral pixel size (channel width) is 1.25 \ang\, corresponding to about 40 km/s.
Upper and lower panels denote the \oii\ emission line at 3726 \ang\ and 3729 \ang, respectively. 
Note that the last map of the upper panels and the first map of the lower panels represent transitional channels and show the contribution of both the receding \oii-3726\ang\ and the approaching \oii-3729\ang\ emission. 
Contour levels are set at $2\times n \times \sigma_\mathrm{noise}$, where $\sigma_\mathrm{noise}=2.0\times10^{-20}$ erg s$^{-1}$cm$^{-2}$\ang$^{-1}$ and $n=1,2,4,8$. Negative contours, shown in grey in the data and in the residuals, are at -2$\sigma_\mathrm{noise}$. The cyan crosses denote the centre of the galaxy.}
\end{figure*}

Our VLT/MUSE data provide for the first time valuable and spatially-resolved integral field observations of the M3 image, which we use to derive the kinematics of \gal\ with unprecedented accuracy. 
Moreover, we use the M1 image, which has higher magnification than M3 and hence higher physical spatial resolution, to test the reliability of the inferred kinematics in the inner parts of the galaxy.
The high-quality of the MUSE data-cube can be appreciated in \autoref{fig:chmaps}, where the blue contours trace the \oii\ emission in the M3 image at different wavelengths (the so-called \vir{channel maps}).
In the \emph{right panels} of \autoref{fig:maps}, we also show moment maps derived by fitting a double Gaussian function to each spatial pixel in the \oii\ data-cube of the M3 image.
The velocity field shows a strong velocity gradient along the major axis that points to a manifest rotation. 
The values of velocity dispersions in the dispersion field are not beam-smearing corrected and are completely dominated by instrumental effects. 
We stress that we do not make use of these moment maps in our kinematic modelling (see \refsec{sec:kinematics}).

We measured the Point Spread Function (PSF) of our observations from the bright star visible in the \emph{left panel} of \autoref{fig:maps}. 
We fitted a 2D Gaussian function to the integrated map of the star and we found a spatial resolution $0.90''\times0.85''$, where the values are the full widths at half maximum (FWHMs) of the best-fit Gaussian function along the right ascension and declination directions, respectively. The PSF, compared to the size of the galaxy in the M3 image, is shown in the \oii\ total intensity map (\autoref{fig:maps}, \emph{right panels}).
We point out that the smearing caused by the PSF makes the small spatial variations due to lensing shear negligible.
The Line Spread Function (LSF), which determines the spectral resolution, of data is 2.3 \ang\ (FWHM).
The contribution of the spectral broadening to the observed dispersion of the line profiles is therefore 0.98 \ang, corresponding to $\sigma_\mathrm{instr}\simeq31$ km/s at the wavelengths correspondent to the redshifted \oii\ doublet.

\section{Stellar mass estimates}
\label{sec:stellarmass}

We measured cumulative stellar mass values within each aperture considered in the kinematic modelling (see \refsec{sec:kinematics}) by fitting the galaxy spectral energy distribution (SED), obtained in the 12 reddest CLASH HST bands, with stellar population models based on \citet{Bruzual&Charlot03} single-burst (SSP) templates. 
We note that, since the global star formation rate of galaxies reaches a maximum at $z=1-2$, the commonly used declining, delayed or truncated star formation histories (SFHs) might not apply to a $z \simeq 1.5$ star-forming galaxy like Sp1149. 
In this case, the fit would typically yield unphysical solutions (e.g., extremely young stellar populations), especially if the rest-frame UV continuum is sampled more finely by the photometry than the rest-frame optical-NIR. 
Here we instead used a rising SFH based on the evolution of the M$_{*}$-SFR relation, using the parametrization of \citet{Sargent+14}. 
This main sequence SFH starts at $z=10$ and has one free parameter, a seed mass which we let vary between $10^5$ and $10^8$ $M_{\odot}$. 
For this model we also assumed an evolving stellar metallicity, computed at each time step from a gas metallicity based on the fundamental mass-metallicity relation of \citet{Mannucci+10} \citep[see][for more details]{Gobat&Hong16}. 
We bracketed it by the metallicity range of the templates (0.005-2.5 $Z_{\odot}$) and interpolated linearly on this grid when combining the SSP spectra. 
We also included extinction by dust, assuming a \citet{Calzetti+10} law with $A_{\text{V}}$=0--4 mag. 
We then convolved the resulting model spectra with the response of the HST filters and rescaled the photometric errors so that the reduced chi-square of the fit be no more than 1. 
We assumed a \citet{Kroupa01} stellar IMF throughout and converted all relations that make use of another type to this one.
The estimated stellar mass values within each aperture are listed in \autoref{tab:table}.

\section{Kinematic models}
\label{sec:kinematics}
We derived the kinematic parameters of \gal\ by using a modified version of \bba\ \citep{DiTeodoro&Fraternali15}, an algorithm that fits 3D tilted-ring models to emission-line data-cubes. 
\bba\ basically produces a disc model and shapes it into a 3D simulated observation by building a line profile in each spatial pixel. 
Instrumental effects are taken into account by convolving the model with the PSF and the LSF of the observations. 
The model-cube is eventually compared pixel-by-pixel to the observed data-cube. 
This 3D methodology allows us to overcome the well-known difficulty of deriving galaxy kinematics from 2D maps in low spatial resolution observations \citep[e.g.,][]{Newman+13}. 
Extensive tests with simulated and local galaxies have shown that \bba\ is very effective in recovering the intrinsic kinematics whenever the disc is resolved with at least 3 independent resolution elements across the major axis \citep[see, in particular, Fig. 7 and Fig. 8 in][]{DiTeodoro&Fraternali15}. 
Moreover, unlike 2D techniques, where the kinematic models may depend on the way velocity and dispersion fields are derived from the data-cube, our approach directly applies on the original dataset and does not require the extraction of any additional kinematic maps.
 
\bba\ builds a kinematic model assuming that a disc galaxy can be decomposed in a certain number of rings, each of which is defined by six parameters: 1) the galaxy centre coordinates $(x_0,y_0)$, 2) the inclination angle $i$ of the galaxy with respect to the plane of the sky, 3) the position angle $\phi$ of the major axis, taken anticlockwise from the North direction, 4) the redshift $z$ of the galaxy, 5) the rotation velocity $\vrot$ and 6) the intrinsic velocity dispersion $\vdisp$ of the gas. All these parameters can in principle vary ring by ring.

\begin{figure}
\label{fig:pvs}
\center
\includegraphics[width=0.5\textwidth]{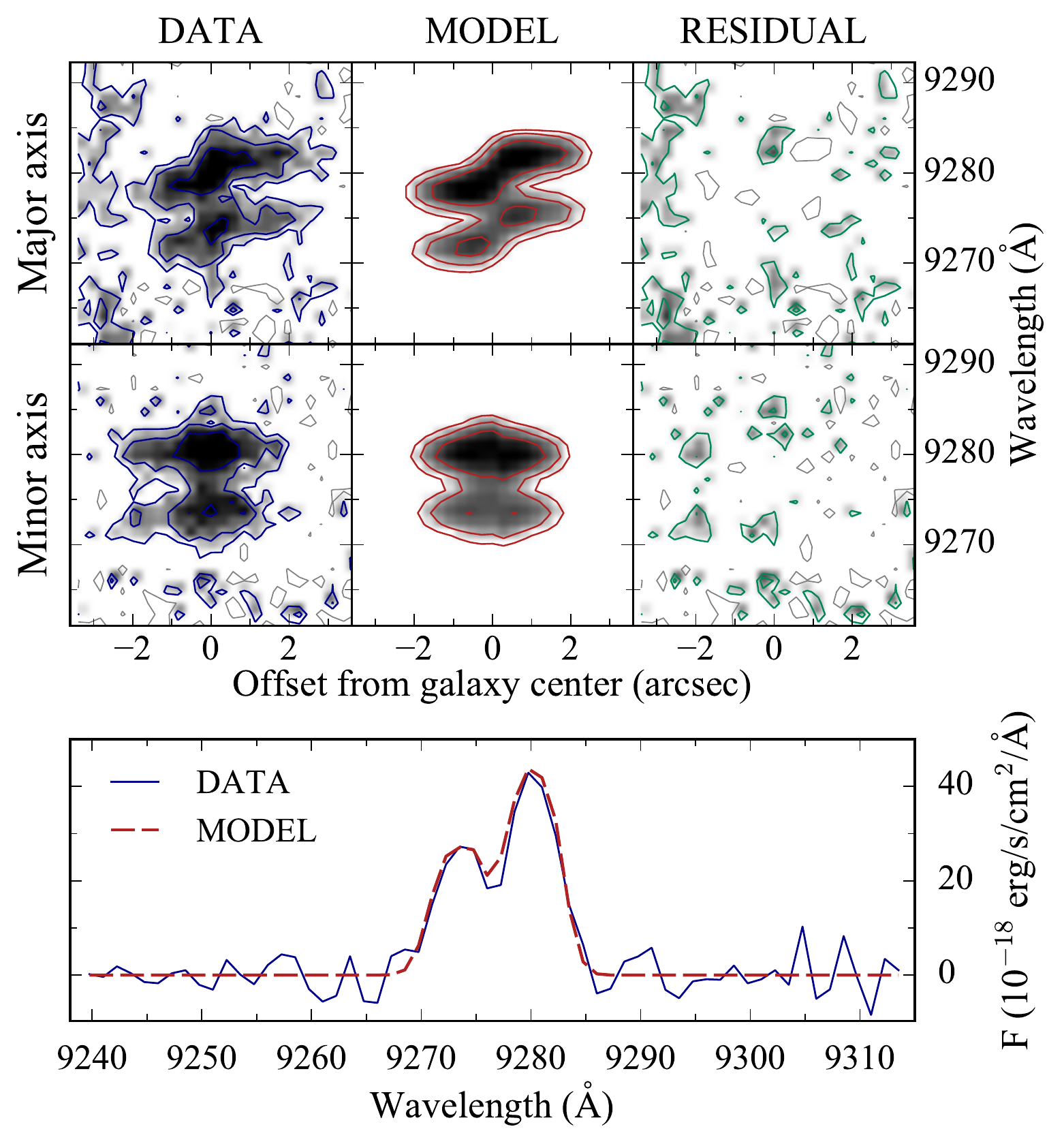}
\caption{\emph{Top:} position-velocity (PV) diagrams along the major axis (\emph{top}, $\phi=227\de$) and the minor axis (\emph{bottom}, $\phi=137\de$) for the M3 image. \emph{Left-hand panels} show the data, \emph{middle panels} the model, \emph{right-hand panels} the residuals (data minus model). Contour levels are the same of \autoref{fig:chmaps}. A pixel in the $y$-axis corresponds to about 40 km/s. The vertical emission strip in the major axis PV plots is due to the continuum of a nearby galaxy.
\emph{Bottom:} comparison between the total \oii\ line profiles of the data (blue) and our best-fit model (red dashed line). 
}
\end{figure}

The standard version of \bba\ works on single emission lines. In order to exploit the entire kinematic information in our data, we updated the algorithm to fit both lines of the \oii\ doublet at once. In our modified version, the line profile $I(\lambda)$ in each spatial pixel is built by adding the contribution of two Gaussian components $G_1$ and $G_2$:

\begin{equation}
\label{eq:2GAU}
\begin{aligned}
I(\lambda) &=  G_1(A_1, m_1, \sigma_1) + G_2(A_2, m_2, \sigma_2) = \\
			     &=  G_1(A_1, m_1, \sigma\phantom{{}_1})  + G_2(\kappa A_1, m_1 + \Delta\lambda, \sigma)
\end{aligned}
\end{equation}

\noindent where $A_\mathrm{n}$, $m_\mathrm{n}$ and $\sigma_\mathrm{n}$ are the amplitude, mean and standard deviation of the $n=1,2$ component. In \autoref{eq:2GAU}, we assume that $G_2$ is scaled by factor $\kappa = A_2/A_1$, has an offset of $\Delta\lambda = m_2 - m_1 $ from $G_1$ and equal standard deviation $\sigma = \sigma_1=\sigma_2$. For the \oii\ doublet, we set $\Delta\lambda = 2.7 \,(1+z)\;\mathrm{\AA}=6.7$ \ang, where 2.7 \ang\ is the wavelength separation at rest of the two lines in the doublet,  and we estimated $\kappa=1.62$ from the total spectrum of \gal\ (see \autoref{fig:pvs}, \emph{bottom panel}). The standard deviation $\sigma$ of the Gaussian functions is a combination of the gas velocity dispersion $\vdisp$, which is a parameter of the model, and the instrumental LSF $\sigma_\mathrm{instr}$.
We tested the updated algorithm with simulated galaxies \citep[following the procedure described in Section 4 of][]{DiTeodoro&Fraternali15} and found that the performance of \bba\ with emission line doublets is fully comparable to the single emission line case.  

Although the algorithm is able to simultaneously fit up to six parameters per ring, this is not recommendable in low-resolution data. 
We decided to fit only the kinematic parameters ($\vrot$ and $\vdisp$), while the other parameters were evaluated \emph{a priori} and then kept fixed in the modelling step. We estimated these parameters as follows:\\

\begin{itemize}
\item \emph{Galaxy centre}. HST observations of MACS1149 are publicly available from various surveys (e.g. HFF, CLASH). We made use of HST combined images from CLASH in F150W/F814W/F225W bands to estimate the galaxy centre. We fitted a S\'ersic profile \citep{Sersic63,Caon+93} to the galaxy surface-brightness by using the software \textsc{imfit} \citep{Erwin+15} and we found $(x_0,y_0)$ = ($\upalpha$=11:49:36.819,  $\updelta$=+22:24:08.80) (J2000), with $1\upsigma$ uncertainties of 0.1$''$. The adopted centre is shown as a cross in the zoom-in panel of \autoref{fig:maps}. We note that the \oii\ total intensity (\emph{top-right} map) well matches the stellar light distribution and that the centre estimated from the HST images coincides with the central peak of the \oii\ emission.\vspace*{0.2cm}

\item \emph{Redshift}. We fitted a 1D double Gaussian function to the \oii\ doublet in the MUSE total spectrum and we measured $z=1.4888\pm0.0011$, a value in good agreement with the previous literature (e.g. $z=1.491$, \citealt{Smith+09}; $z=1.489$, \citealt{Kelly+15}).\vspace*{0.2cm}

\item \emph{Inclination}. The galaxy appears to be nearly face-on, which makes it challenging to estimate its actual inclination. \citet{Yuan+11} and \citet{Livermore+15} quote a $i=45\de$, coming from a kinematic fit of the H$\upalpha$ velocity field of the M1 image. 
The M3 HST image and the M3 \oii\ intensity map suggest however a slightly lower inclination angle.
Our \textsc{imfit} fit on the M3 HST image returns an $i_\mathrm{HST}=(37\pm6)\de$. 
We also used an algorithm, implemented in \bba, that fits a PSF-convoluted model map to the observed intensity map and we found  $i_\mathrm{OII}=(32\pm7)\de$. 
We eventually decided to set $i=35\de$ as our best guess in the following kinematic modelling. However, given the wide range of acceptable values and having no further information to constrain the real inclination angle, we decided to calculate the uncertainties on rotation velocity and velocity dispersion by fitting two additional models: a nearly face-on disc at $25\de$ and, for consistency with former studies, a more inclined disc at $45\de$ (see below for a discussion on the error estimate).
\vspace*{0.2cm}

\item \emph{Position angle}. We extracted a velocity field by fitting a 1D double Gaussian function with fixed $\Delta\lambda = 6.7$ \ang\ to each spatial pixel of the \oii\ MUSE data-cube (\autoref{fig:maps}, \emph{right panels}). 
The characteristic velocity $V(x,y)$ along the line-of-sight at a given position $(x,y)$ was calculated from the offset of the first Gaussian function from the redshifted \oii-$\uplambda3726\mathrm{\AA}$ line, such that $V(x,y)=c\,(m-\lambda_z)/\lambda_z$, where $m$ is the mean of the Gaussian, $c$ is the speed of light and $\lambda_z = \lambda_0 \,(1+z) = 9274\,\mathrm{\AA}$. 
We estimated the position angle $\phi$ by finding the straight line passing through the galaxy centre and along which the gradient on the velocity field is maximum. This procedure is directly provided by \bba. We found $\phi=(227\pm6)\de$. The adopted position angle is shown as a straight line on the \gal\ velocity field in \autoref{fig:maps}.
\end{itemize}
\vspace*{0.3cm}

During the fitting procedure, the MUSE data-cube was masked through a S/N threshold cut equal to 2.5, i.e. we blanked all those regions where the \oii\ flux is lower than 2.5 times the noise level on a channel-by-channel basis. We calculated the noise level as the root mean square (RMS) in five different emission-free regions of each channel of the data-cube. 
Estimating the RMS at each wavelength separately prevents a particularly noisy channel from invalidating the goodness of the fit.
We stress however that the noise level in our MUSE data is consistently constant over the wavelength range covered by the \oii\ lines, with an average value $\sigma_\mathrm{noise}=2.0\times10^{-20}$ erg s$^{-1}$cm$^{-2}$\ang$^{-1}$. 
In addition to the S/N cut, we manually masked a 1.3$''$ circular region centred at ($\upalpha$=11:49:36.970, $\updelta$=+22:24:10.83), where we detected the continuum emission of the foreground galaxy located North-West from \gal\ M3 image (see HST image in the \emph{right panel} of \autoref{fig:maps}).

We built kinematic models by using ring widths of 0.4$''$, a value that is twice the pixel size and about half the FWHM of the PSF. 
Models were normalized azimuthally, i.e. we calculate and impose a constant gas surface density along each ring. We also checked that a fit with the local normalization used in \citet{DiTeodoro+16} returned consistent best-fit values for $\vrot$ and $\vdisp$.
A $\cos(\theta)$ weighting function, being $\theta$ the azimuthal angle, was applied to the residual calculation in order to give more importance to the pixels lying closer to the major axis ($\theta$=0).
Errors in \bba\ are calculated through 1000 Monte Carlo realizations of the model, where the  parameters are randomly drawn from a Gaussian distribution centred on the best-fit values \citep[see][for further details]{DiTeodoro&Fraternali15}. 
In this process, all the parameters are varied, including those kept fixed during the fit. 
The largest uncertainties on the kinematic parameters, in particular on the rotation velocity, comes from the inclination angle: \gal\ is seen nearly face-on and small variations of the inclination produce large changes in the derived rotation velocity (the observed component along the line-of-sight is $V_\mathrm{LOS} = \vrot \sin \, i$).
As a conservative approach, we extended the \bba's error budget by fitting a more face-on model ($i=25\de$) and a more inclined model ($i=45\de$). 
The best-fit values at $i=25\de$ and $i=45\de$ were used to quantify upper and lower limits to the rotation velocities and the velocity dispersions of \gal.
In this way, we can account for an effective variation of the inclination $\Delta i = \pm10\de$ from our fiducial model at $i=35\de$, a range sufficiently large to embody also possible lensing uncertainties.

The kinematic analysis was performed on the M3 image of the galaxy with no prior lensing reconstruction. 
Rotation curve and velocity dispersion profile were derived as a function of the angular distance in the image-plane from the galaxy centre, which was afterwards traced back to source-plane radii using the MLV-G12F median lensing model by \citep{Grillo+16}. 
We stress however that our best-fit values of rotation velocity and intrinsic velocity dispersion are not dependent on the assumed lensing model, which instead exclusively affects the conversion between a radius in the image plane to a radius in the source plane.

\begin{figure*}
\label{fig:param}
\center
\includegraphics[width=\textwidth]{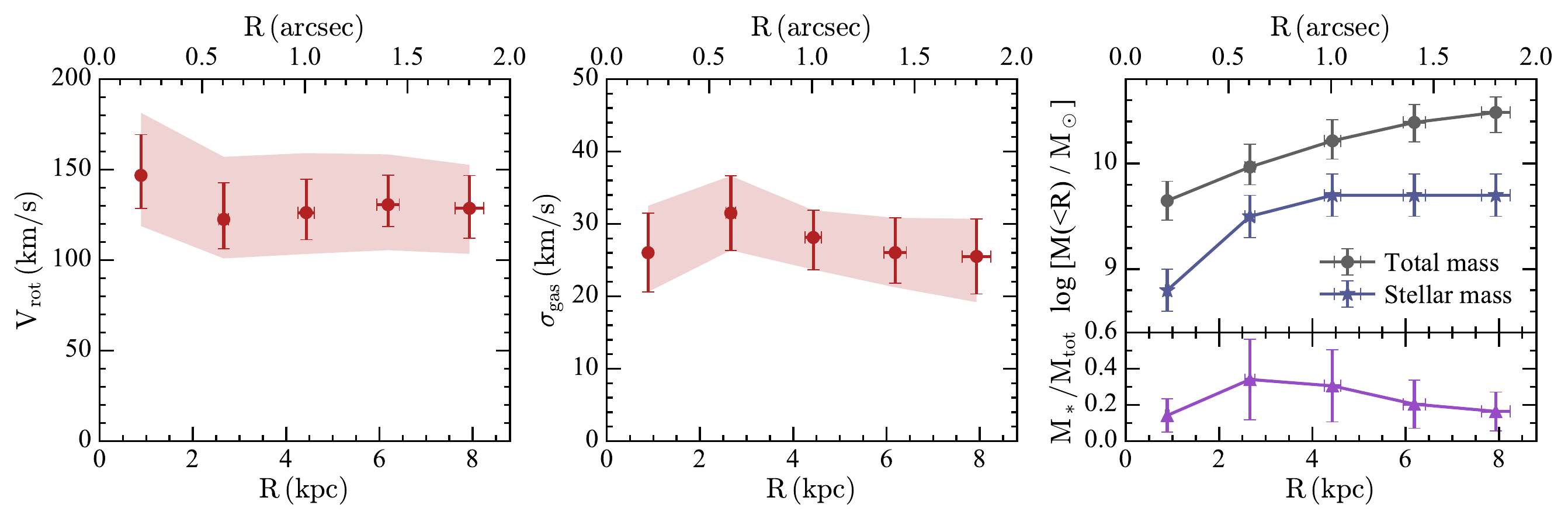}
\caption{Rotation curve (\emph{left}), velocity dispersion (\emph{middle}) and stellar/dynamical masses (\emph{right}) for the M3 image of \gal. In the \emph{left} and \emph{middle panels}, the red points with errorbars are relative to the \bba\ best-fit model with $i=35\de$, while the shadowed strip represents a wider confidence region estimated by using best-fit models at $i=25\de$ and $i=45\de$ (see text). In the \emph{right panel}, blue stars, gray points and purple triangles denote the cumulative stellar masses (see \refsec{sec:stellarmass}), total dynamical masses and stellar-to-total mass fractions (see \refsec{sec:results}), within radius $R$, respectively.
Radii in arcsec are in the image plane, radii in kpc are in the source plane. Magnification factors to convert angular to linear dimensions are taken from \citet{Grillo+16} (see \autoref{tab:table}).}
\end{figure*}

\section{Results and discussion}
\label{sec:results}

\autoref{fig:chmaps} compares the M3 MUSE data-cube to our best model-cube at $i=35\de$ in the whole wavelength range where the \oii\ doublet is detected. The upper panels refer to the \oii-3726\ang\ line, the lower panels to the \oii-3729\ang\ line. 
The data is shown in the top row (blue contours), the model in the middle row (red contours) and the data-minus-model residuals in the bottom row (green contours).
Our model is completely symmetric both in surface density and kinematics. 
The model nicely traces the kinematic behaviour of the galaxy and, in particular, the channel maps at the extremest wavelengths ($\lambda=9269.76$ \ang\ and $\lambda=9283.51$ \ang), which carry information on the maximum rotation velocity, are well reproduced. 
The strong residuals visible in the top and bottom left corners of all channel maps come from the continuum emission of two nearby galaxies, members of the cluster (see \autoref{fig:maps}). 
We also extracted position-velocity diagrams (PVs), namely slices through the spectral dimension of the data-cube, to check that the model is consistent with the observations. 
In \autoref{fig:pvs} (\emph{top panels}), we show PVs taken along the major (\emph{top}) and minor (\emph{bottom}) axes of the galaxy. 
We plot PVs of data, model and residuals in the left, middle and right columns, respectively. 
Residuals show that the model replicates the bulk of the \oii\ emission in the galaxy disc. 
The vertical strip in the left part of the major axis and the residual emission in the left side of the minor axis are due to the continuum from the top-left galaxy and the bottom-left galaxies visible in \autoref{fig:chmaps}, respectively.
Some residual emission emerges at high velocities in the very central regions both on the major and minor axes. 
These features can be either a high-velocity \oii\ emission component or the galaxy continuum, which has not been subtracted in this work and might be stronger in the galaxy center.  
In the first case, the high-velocity central emission can be reproduced either by a very high gas surface density and rotation velocity ($\sim 250$ km/s) in the inner region of the galaxy or by non-circular motions \citep[outflows, e.g.][]{Bradshaw+13}.
In the \emph{bottom panel} of \autoref{fig:pvs}, we finally show that the model well matches the overall kinematic properties of the galaxy by comparing the 1D global spectra of data (blue) and best-fit model (red dashed).

The derived rotation curve and velocity dispersion profile are reported in \autoref{fig:param}. 
Radii in arcsec units in these plots are measured on the image-plane and then converted to source-plane physical radii by using the magnification factors from \citet{Grillo+16} listed in \autoref{tab:table}. 
Uncertainties in $R$ are propagated from the statistical errors of the lensing model: best-fit uncertainties on the magnification factors are always lower than 0.2, thus we used a conservative value of 0.2 to compute errors on delensed radii.
In both plots we show the best-fit values of the model with $i=35\de$ (red points with errorbars) and a \vir{safety} strip (red region) where we include the uncertainties on the inclination angle in a very conservative way. 
The lower (upper) boundaries of this region are calculated as the minimum (maximum) between the best-fit values of the $i=35\de$ model minus (plus) the error and the values obtained from the $i=25\de$ and $i=45\de$ models. 
The rotation curve (\emph{left panel} of \autoref{fig:param}) of \gal\ peaks at 148 km/s around 1 kpc from the galaxy centre, it shows a slight decline and then flattens to an average value $V_\mathrm{flat}\simeq 128$ km/s, where the average is calculated over all the points but the innermost one. 
The actual value of $V_\mathrm{flat}$ is however strongly affected by the inclination uncertainty and it could raise to 157 km/s for $i=25\de$ or drop to 109 km/s for $i=45\de$. 
As a benchmark, the local stellar-mass Tully-Fisher relation for a galaxy with $\log(M_*/M_\odot) = 9.7$ predicts a rotation velocity $V_\mathrm{flat}\simeq125$ km/s \citep[e.g.][]{McGaugh&Schombert15}. 
This velocity is fully compatible with the value we found for our fiducial model at $i=35\de$, within the errors.
The flat shape of the outer rotation curve is similar to that observed in low-$z$ spiral galaxies and conflicts with recent claims of declining rotation curves at $z\sim1-2$ \citep{Genzel+17, Lang+17}.
The central peak of the rotation velocity implies a very steep rise of the curve within a 0.2$''$ radius and suggests a strong concentration of matter in the inner kpc \citep[e.g.][]{Sofue+99}, consistent with the size of the bulge-like structure in the galaxy centre (see HST images in \autoref{fig:maps}). 

The velocity dispersion (\emph{middle panel} of \autoref{fig:param}) is nearly constant across the galaxy disc, with an average value of $27\pm5$ km/s. 
We note that the uncertainties on the measured velocity dispersions do not depend on the inclination angle assumed and the limits of the confidence strip are mostly set by the error obtained in the fit at $i=35\de$. 
Our value of 27 km/s is in good agreement with the ionized gas velocity dispersions of local star-forming galaxies \citep[e.g.][]{Andersen+06,Epinat+10} and recent 3D determinations at $1\lesssim z \lesssim2$ \citep{DiTeodoro+16, Mason+16}, while it is generally lower than 2D measurements at similar redshifts: in particular, our velocity dispersion is about half of the average values found by \citet{Stott+16} at $z\simeq0.9$, by \citet{Epinat+12} at $z\simeq1.3$ and by \citet{Wisnioski+15} at $z\simeq2$ and a factor 4-5 lower than the dispersions quoted by \citet{ForsterSchreiber09} at $z\simeq2$. 
We stress that the combination of lensing magnification, low inclination of the galaxy and 3D modelling allows us to minimize the impact of instrumental and projection effects and to robustly derive the intrinsic velocity dispersion of the gas.
Since the velocity dispersion is a measure of non-ordered random motions (turbulence), we conclude that \gal's disc is not significantly more turbulent than today spiral galaxies with similar SFR. 
\gal's disc may nonetheless be thicker than local discs because of the shallower potential of the galaxy.
The low velocity dispersion is also consistent with the existence and the survival of the prominent spiral arm features in \gal\ \citep{Elmegreen&Elmegreen14, Yuan+17}.

We can roughly estimate the stellar-to-total mass fractions $M_*/M_\mathrm{tot}$ at different radii.
The total dynamical mass in a spherically symmetric system can be inferred from the rotation curve: the mass contained inside a radius $R$ is $M_\mathrm{tot} (<R) = R V^2_\mathrm{rot}(R)/G$, where $\vrot(R)$ is the rotation velocity at $R$ and $G$ is the gravitational constant. 
Given the low velocity dispersion, we can neglect the contribution of random motions to the dynamical support of the galaxy \citep[asymmetric drift,][]{Oort65}. 
We calculated dynamical masses inside each radius and we compared them to the stellar masses derived in \refsec{sec:stellarmass}. In the \emph{right panel} of \autoref{fig:param} we show the cumulative trends with radius of the derived stellar masses (blue stars), total dynamical masses (gray points) and stellar-to-total mass fractions (purple triangles).
The average stellar-to-total mass fraction is $ \langle M_*/M_\mathrm{tot} \rangle \eqsim 0.2\pm0.1$, meaning that gas and dark matter contribute to about 80\% of the total mass budget.
This value is on the low stellar-to-total mass fraction side but still consistent with those of recent studies at similar redshifts \citep[e.g.,][]{Stott+16,Wuyts+16,Pelliccia+17} and some works at low redshift \citep[e.g.,][]{Bershady+11,Martinsson+13}. 
In \autoref{tab:table} we summarize all the relevant parameters in this work, including used magnification factors, derived rotation velocities, velocity dispersions, stellar masses and dynamical masses.

\begin{table}
\caption{Parameters for \gal\ at different radii derived in this work from the M3 image.  (1) Radius in arcsec in the image-plane. (2) Radius in kpc in the source-plane, using the MLV-G12F model of \citet{Grillo+16}. (3) Average flux magnification factor within $R$ from the same lensing model. Typical statistical error is < 0.2. (4)-(5) Derived rotation velocity and velocity dispersion for our best model at $i=35\de$. Quoted errors take into account a $\Delta i=\pm10\de$ (see text). (6) Stellar mass within $R$, derived as described in \refsec{sec:stellarmass}. Typical error is 0.2 dex. (7) Dynamical mass within $R$, derived from the rotation velocity.}
\label{tab:table} 
\centering
\def\arraystretch{1.3}
\begin{tabular}{ccccccr}
\hline\hline\noalign{\vspace{5pt}}
R & R & $\mu$ & $\vrot$ & $\vdisp$ & \multirow{ 2}{*}{$\log \frac{M_*}{M_\odot} $} & \multirow{ 2}{*}{ $\log\frac{M_\mathrm{tot}}{M_\odot}$} \\
($''$) & (kpc) &  & (km/s) & (km/s) &  &  \\
(1) & (2) & (3) & (4) & (5) & (6) & (7) \\
\noalign{\smallskip}
\hline\noalign{\vspace{5pt}}
0.2  & 0.9 & 3.8 & 148$\pmo{35}{27}$  &	 26$\pmo{6}{5}$   & 8.8  & 9.65$\pmo{0.19}{0.18}$  \\
0.6  & 2.7 & 3.8 & 123$\pmo{34}{22}$  &  31$\pmo{5}{5}$   & 9.5  & 9.98$\pmo{0.21}{0.17}$  \\
1.0  & 4.4 & 3.8 & 127$\pmo{32}{23}$  &  28$\pmo{4}{4}$   & 9.7  & 10.22$\pmo{0.20}{0.17}$ \\
1.4  & 6.2 & 3.8 & 131$\pmo{28}{25}$  &  27$\pmo{5}{5}$   & 9.7  & 10.40$\pmo{0.15}{0.16}$  \\
1.8  & 7.9 & 3.9 & 129$\pmo{24}{23}$  &  25$\pmo{5}{6}$   & 9.7  & 10.49$\pmo{0.14}{0.16}$  \\
\noalign{\vspace{2pt}}\hline
\noalign{\vspace{10pt}}
\end{tabular}
\end{table}

\begin{figure}
\label{fig:M1M3comp}
\center
\includegraphics[width=0.4\textwidth]{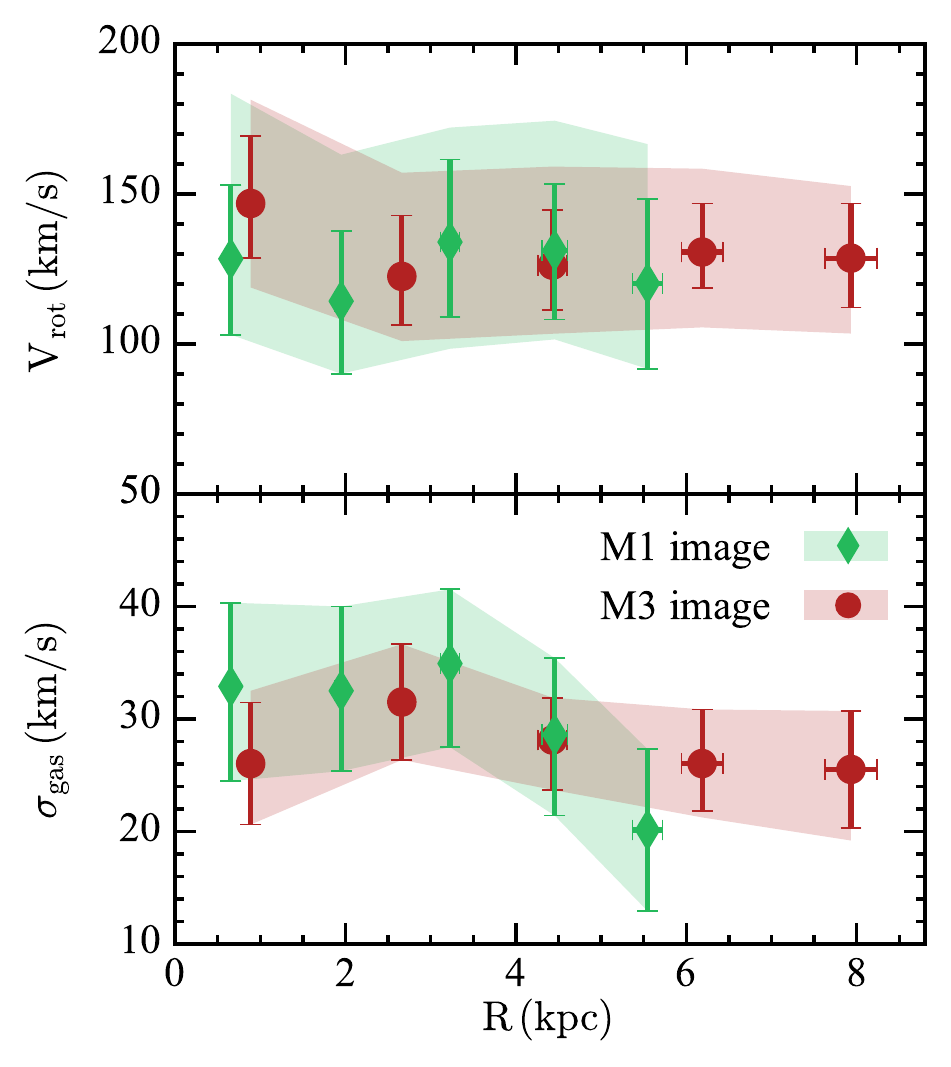}
\caption{Comparison between the best-fit rotation curve (\emph{top panel}) and velocity dispersion profile (\emph{bottom panel}) derived from image M1 (green diamond) and image M3 (red points). Shaded areas are as in \autoref{fig:param}.}
\end{figure}

As a final sanity check, we repeated the same kinematic analysis on the M1 image in our MUSE data. 
M1 is more magnified \citep[$\mu\sim8$,][]{Grillo+16} than M3 but heavily sheared, especially in the external regions of the disc where the gravitational lensing effects due to the foreground elliptical cluster member (see \autoref{fig:maps}) are not negligible.
However, M1 can be used to test the kinematics of the inner regions with a slightly higher linear resolution (i.e.\ 2.7 kpc for M1 vs 3.9 kpc for M3).
For the kinematic fit, we fixed redshift, inclination and position angles to the values found for M3.
The galaxy centre was fixed to the brightest pixel in the CLASH HST composite image. All other parameters were left as described in  \refsec{sec:kinematics}. 
In \autoref{fig:M1M3comp} we show the best-fit rotation curve (\emph{top panel}) and velocity dispersion profile (\emph{bottom panel}) for M1 (green diamonds) compared to M3 (red points). 
The M1 curves are truncated at $R < 6$ kpc because at larger radii the fit can not converge due to the significant distortion.
Despite the fact that we do not take into account gravitational lensing, the inner kinematics derived from M1 is in good agreement with that derived from M3. 
In particular, the higher spatial resolution supports the scenario of a quickly rising rotation curve, followed by a slight decline and a flat part.
Velocity dispersions are consistently low, i.e.\ around 30 km/s, at all radii.
The higher distortion of M1 mainly reflects in higher uncertainties.

\section{Comparison with former studies}
\label{sec:comp}
The kinematics of the galaxy \gal\ has been studied in three previous papers, namely \citet{Yuan+11}, \citet{Livermore+15} and \citet{Mason+16}.
\citeauthor{Yuan+11} and \citeauthor{Livermore+15} used AO observations with the Keck/OSIRIS IFU of the \ha\ and \nii $\uplambda\uplambda6548-6583\AA$ emission lines in the M1 image, with total exposure time of 4.75 hours.
\citeauthor{Mason+16} used seeing-limited (median seeing $\sim 0.6 ''$) VLT/KMOS IFU observations of the \oiii$-\uplambda5007\AA$ emission line in the M1 and the M3 images with 2.25 hours of integration time on each source.
The kinematics derived in these three studies and that presented in this work do not agree. 
The discrepancies can not be ascribed to the use of different emission features because \ha, \nii, \oii\ and \oiii\ lines arise from the warm ionized gas and are reasonably expected to trace the same kinematics.

The paper of \citet{Yuan+11} focused on the metallicity gradient in \gal\ and a proper kinematic modelling is beyond their goals. They estimated a value for the rotation velocity by fitting a model velocity field \citep{Jones+10b} to the observed velocity field of the M1 image after reconstruction on the source plane based on the lens model of \cite{Smith+09}. 
Beam smearing is not taken into account. 
They quote a rotation velocity $\vrot\sin(i) = 150 \pm 15$ km/s. 
An inclination angle of $35\de$ would lead to $\vrot \simeq 260$ km/s, a value which is hardly compatible with our rotation curve. 
It is difficult to assess the goodness of \citeauthor{Yuan+11}'s kinematics because their model is not shown in the paper. 
A direct comparison with our kinematic estimate is also difficult because they do not specify at which radius the rotation velocity is measured.
Our data however seem to rule out the possibility of a value of 150 km/s for the line-of-sight velocity: in such case, the combined effect of rotation velocity, velocity dispersion and instrumental smearing would have made the galaxy to spread its emission in channels with velocities well beyond those detected in our data-cube (for example, in the first channel of \autoref{fig:chmaps}).

\citet{Livermore+15} included the dataset from \citet{Yuan+11} in a broader study on the kinematics of a sample of lensed galaxies.
Similarly to \citeauthor{Yuan+11}, they fitted the delensed 2D kinematic velocity map with an arctangent model of the rotation curve \citep{Courteau97} without any correction for beam smearing and found a rotation velocity $V_{2.2} = 59 \pm 3$  km/s at $R_{2.2} = 1.32 $ kpc for an inclination $i=45\de$, where $R_{2.2}$ is the radius at 2.2 times the scale length of an exponential disc.  
Although MUSE data have lower spatial resolution than OSIRIS data, our rotation curve seems to exclude such a low rotation velocity at 1.32 kpc (\autoref{fig:param}, \autoref{fig:M1M3comp}), even after correcting for the different inclination ($V_{2.2} \simeq 73$ km/s for $i=35\de$) . 
Since \citeauthor{Livermore+15} do not report their asymptotic velocity nor turnover radius of the best-fit arctangent function, we can not compare the values of the flat part of the rotation curve.
To constrain the velocity dispersion, \citeauthor{Livermore+15} fitted an exponential model to the velocity dispersion map, which they derived applying a correction for beam smearing and spectral broadening.
They quote a luminosity-weighted average of $\upsigma = 50 \pm 10$ km/s. This value is however strongly affected by the two inner points (see their Fig. 2), which appear to be dominated by residual beam smearing.
Finally, we stress that the \citeauthor{Livermore+15}'s kinematic model does not satisfactorily reproduce their data (see the residual velocity field in their Fig. 1).
This is likely due to a combination of the low S/N AO data, some beam smearing and the uncertainties related to the lensing modelling.

\citet{Mason+16} made use of the 3D fitting code GalPaK$^\mathrm{3D}$ \citep{Bouche+15} on KMOS \oiii$-\uplambda5007\AA$ data. Like \bba, this code fits 3D models to data-cubes and properly accounts for beam smearing.
The main difference between this code and \bba\ is that, in GalPaK$^\mathrm{3D}$, the rotation curve follows a preselected functional form  and the velocity dispersion is always assumed to be constant across the disk. 
\citeauthor{Mason+16} have observations of both M1 and M3 images. 
They could not fully model the M3 image because it is unresolved. 
M1 image is poorly detected with S/N$\sim$2 in their data and the KMOS field-of-view is not sufficiently large to include the entire galaxy disc.
They modelled the M1 image without any reconstruction on the source plane and found an asymptotic rotation velocity of $227\pm32$ km/s, an inclination of $12\de\pm2\de$ and a velocity dispersion of $15\pm7$ km/s. 
Their 95\% confidence interval for the asymptotic velocity ranges between 179 km/s and 293 km/s, for the inclination between $8\de$ and $17\de$ (Mason private communication).
The high rotation velocity is due to the low inclination. 
If we rescale their interval to our fiducial estimate of 35$\de$ we obtain a velocity in the range $65-106$ km/s, which is marginally consistent with our values.
We finally note that \citeauthor{Mason+16} do not use this source in their analysis of the full sample because of the large uncertainties due to the low S/N and the lack of correction for the high magnification.

\section{Conclusions}
\label{sec:summary}

In this paper we combined the power of VLT/MUSE observations, the gravitational magnification and a state-of-the-art 3D technique to investigate the kinematic properties of the galaxy (\gal) hosting the SN Refsdal. 
\gal, located at $z=1.49$, is multiply imaged and highly magnified by the cluster MACS J1149.5+2223 at $z=0.542$. 
Given its stellar mass ($\log  M_*/M_\odot = 9.7\pm0.2$) and its low star-formation rate ($\mathrm{SFR}=1-6 \, \moyr$), \gal\ looks like how we believe a local spiral galaxy used to be 9-10 Gyr ago.

We used the \oii\ emission doublet, detected in our MUSE data, to model the magnified ($\mu\simeq4$) and almost undistorted M3 image and derive the kinematics of \gal\ directly from the image-plane.
The kinematic modelling was performed by using an updated version of \bba, a code that fits 3D tilted-ring models to emission-line data-cubes. 
Three-dimensional fitting techniques naturally break the degeneracy between rotation velocity and velocity dispersion in low spatial resolution data and return reliable intrinsic kinematics.
The shape of \gal\ rotation curve resembles that of local spiral galaxies: a steeply rising part in the inner kpc followed by a flat part out to large radii. 
Since the galaxy is nearly face-on, the actual value of $\vflat$ strongly depends on the inclination angle. Our fiducial model at $i=35\de$ returns a $\vflat = 128 \pmo{29}{19}$ km/s where the errors account for an inclination uncertainty of $\pm10\de$. 
Despite the large uncertainties, \gal\ appears to lie in the stellar-mass Tully-Fisher relation of local ($z=0$) galaxies.
The velocity dispersion profile is almost flat, with an average value $\vdisp=27\pm5$ km/s. 
Contrarily to the rotation velocity, the velocity dispersion is well constrained because it does not depend on the inclination uncertainty. This value is a factor 2-4 lower than what found in former studies at $1<z<2$ and in good agreement with ionized gas velocity dispersions measured in local spiral galaxies. 
By modelling the HST multi-color photometry, we estimated aperture stellar masses and obtained a value of roughly 20\% for the stellar-to-total (dynamical) mass fraction.

Unlike previous claims of highly turbulent discs at $z>1$, \gal\ appears to be an ordinarily regular spiral galaxy: a well-settled and rotation-dominated disc with $\vsigmaratio\gtrsim5$, where the ionized gas is mildly turbulent and orbits about the galaxy centre following a flat rotation curve. 

\section*{Acknowledgements}

We thank C. Mason and T. Yuan for providing their data and models. 
This work made use of I\textsc{mfit} and \bba\ softwares.
E.d.T. acknowledges the support of the Australian Research Council through grant DP160100723. C.G. acknowledges support by VILLUM FONDEN Young Investigator Programme through grant no.\ 10123. F.F. acknowledges hospitality at the European Southern Observatory in Garching as a scientific visitor. A.M., P.R., G.B.C. and M.L. acknowledge financial support from PRIN-INAF 2014 1.05.01.94.02.

\bibliography{biblio}

\end{document}